# Shear loading of costal cartilage

Damien Subit, Jason Forman

Center for Applied Biomechanics, University of Virginia, USA

## Abstract

A series of tests were performed on a single post-mortem human subject at various length scales. First, tabletop tests were performed. Next, the ribs and intercostal muscles were tested with the view to characterize the load transfer between the ribs. Finally, the costal cartilage was tested under shear loading, as it plays an important in the transfer of the load between the ribs and the sternum. This paper reports the results of dynamic shear loading tests performed on three samples of costal cartilage harvested from a single post-mortem human subject, as well as the quantification of the effective Young's modulus estimated from the amount of cartilage calcification.

## Introduction

Computational injury biomechanics is a very active field of research in the automotive industry with the development of computational models of the whole human body (Global Human Body Models Consortium (GHBMC), Total Human Model for Safety (THUMS), HUman Models for Safety (HUMOS)) that are used to assess the performance of current safety systems and design new countermeasures. The current challenge with these models is the assessment of their biofidelity that relies on results of impact tests performed on Post-Mortem Human Subjects (PMHS): the ribs for instance will be evaluated based on data obtained from a set of PMHS, while the global response of the thorax will be evaluated from a different set of PMHS. While this approach is justified because of the limited experimental data available, it leads to evaluate a single computational model against partial data obtained from multiple PMHS. It is unclear whether this approach leads to a realistic model of the human body or to a *lusus naturae*. In an effort to elucidate this question, a series of tests were performed on a single PMHS torso at various length scales. First, tabletop tests where the torso was loaded with a belt were performed (Subit and Salzar, 2014; Salzar and Subit, 2014). Next, component level tests were performed on the ribs (Subit *et al.*, 2013) and the intercostal muscles (Hamzah *et al.*, 2013). It was demonstrated that (1) the costal cartilage contributed to the transmission of the load between the ribs and the sternum (Murakami *et al.*, 2006), (2) the effective elastic modulus of the costal cartilage increased with the amount of calcification (Forman, 2009; Forman *et al.*, 2010), and (3) the amount of calcification increased with age (McCormik, 1980). Therefore, the objective of this paper was (1) to report the results of dynamic shear tests performed on the costal cartilage, and (2) to quantify the amount of calcification in the costal cartilage samples of this specific PMHS to ultimately estimate their compliance.

## Materials and Methods

This section describes the rationale behind the choice of test conditions, the preparation of the samples, the acquisition of X-ray images prior to testing, and the test apparatus and methodology.

### *Mechanical loading of the costal cartilage under belt loading*

Costal cartilage segments were tested in a loading condition similar to cantilevered bending. This loading condition was chosen to simulate the loading of the cartilage that may occur when the sternum is displaced in the posterior direction towards the spine (e.g., from loading by a seatbelt). Ali *et al.* (2005) studied the deformation of the ribcage under quasi-static anterior loading of the whole chest using computed tomography (CT) imaging of cadavers (Figure 1). That study found that for a 50 kg, 168 cm tall adult male, displacing the sternum posteriorly by 53 mm with a "belt-like" loader resulted in 32 mm and 40 mm of posterior displacement of the right and left 4[th] rib costo-chondral junctions, respectively. This corresponds to posterior displacements of the sternum relative to the CC junctions of 21 mm and 13 mm, respectively. In contrast, the right and left 4[th] rib CC junctions displaced in the lateral direction only 1 mm and 3 mm, respectively (Figure 1). This study indicates that the cantilever bending of the costal cartilage is representative of the loading applied by a belt.

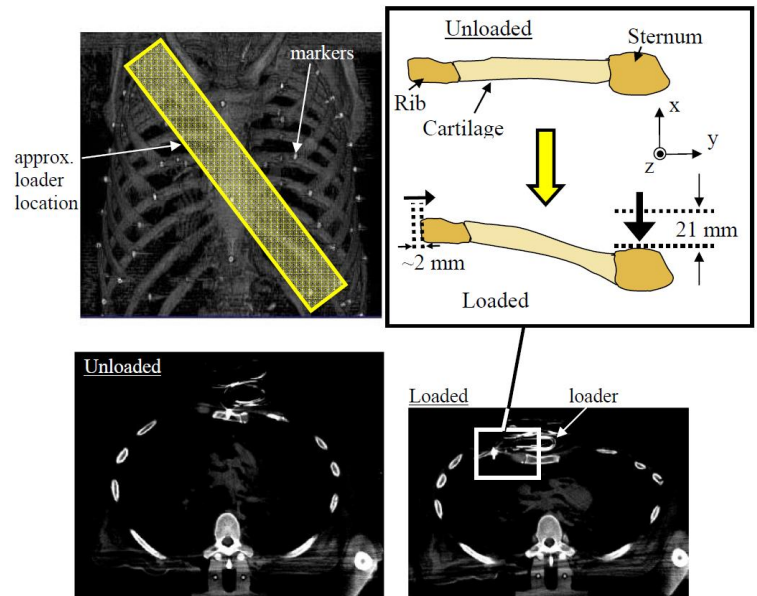

Figure 1: Illustration of the investigation of ribcage deformation using CT scans (Ali et al. 2005). Top Left: CT scan (anterior view) showing the location of radio-opaque markers installed in the ribcage of a cadaver thorax, and the approximate location of a belt-like loader. Bottom: Transverse CT scan slices from the superior-inferior location of the 4th rib illustrating the ribcage cross-sectional geometry before (unloaded) and after (loaded) displacing the sternum posteriorly. Top Right: Illustration of the typical anterior-posterior (x-axis) and lateral (y-axis) displacements of the sternum relative to the costo-chondral junction.

## *Anatomical description of the samples*

Three samples of costal cartilage were extracted from the right side of the rib cage from a single male PMHS (70 year, 160 cm, 60 kg, internal PMHS id: 508). The PMHS was obtained and treated in accordance with the ethical guidelines established by the United States National Highway Traffic Safety Administration, and all testing and handling procedures were reviewed and approved by the University of Virginia Institutional Review Board. The subject was confirmed to be free of infectious diseases (HIV, Hepatitis A and C) prior to any work. The tissue was extracted from the right side of the rib cage and tested using the methods described in Forman *et al.* (2010). The intercostal muscles were cut from the distal end of the costal cartilage (CC), to about 30 mm posterior to the costal cartilage junction (CCJ) (Figure 2). Based on the length of the CC, only the CC from rib level 4 to 7 were considered as appropriate candidates for the tests, as for the other levels, the CC were too short. Further observation of the samples revealed that the CC from rib 5 was fractured at the CCJ as a result of the tabletop tests (Subit et Salzar, 2014), and therefore was too short to be tested. Finally, the CC from rib levels 4, 6 and 7 were prepared for tests. The sample of costal cartilage at the level of rib 4 (CC 4) consisted of a portion of the sternum, the entire length of the cartilage, and about 3 cm of attached rib. As for the samples CC 6 and CC 7, although they were long enough to be tested, only the CC itself could be extracted because of fractures at the costo-chondral and costo-sternal junctions (Appendix 1). All specimens were tested with the perichondrium intact, but CC 4 was the only sample that was similar to that tested in Forman *et al.* (2010).

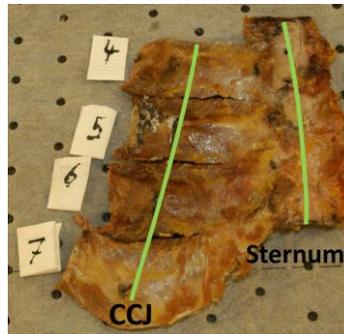
Figure 2: Anterior view of the costal cartilage from the right side of the ribcage, from rib level 4 down to 7.

## *Preparation of the samples for the shear tests*

Prior to potting, the soft tissues were carefully removed from the CC surface without damaging the perichondrium. The sternum and rib ends of the segments were potted in blocks of Fast Cast® (Goldenwest Manufacturing, Inc.) casting resin (approx.. 5 cm × 5 cm × 2.5 cm deep, Figure 3). The specimens were oriented such that the anterior surface of the sternum and the superior-inferior axis of the sternum were perpendicular to the axis of loading. After potting, the specimens were scanned with a clinical computed tomography (CT) scanner (slice thickness of 0.625 mm, in-plane resolution of 0.1875 mm/pixel).

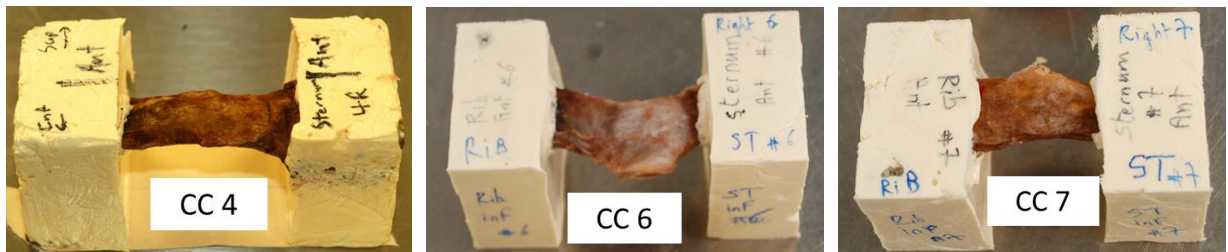
Figure 3: Anterior view of the potted specimen (the sternum block id on the right hand side).

## *Test apparatus and instrumentation*

For testing, the potting blocks were mounted in a test rig powered by an Instron material test machine (model 8500). The test rig was designed to displace the sternum end of the segment in the posterior direction, holding all other degrees of freedom for both potting blocks fixed (Figure 4). All specimens were tested with a dynamic sternum displacement rate of 400 mm/s past the point of specimen failure. The rate of 400 mm/s is consistent with previous studies (Forman *et al.*, 2010; Forman and Kent, 2011), and is based on the cartilage deformation rate estimated to occur during a 48 km/h frontal collision with a mid-sized male restrained by a three-point seatbelt and an airbag (Forman *et al.*, 2010).

Forces and moments in all directions were measured by a six-axis load cell (R.A. Denton Model 5024 J) attached to the rib end of the specimens (Figure 4 and Figure 5). The displacement of the piston was measured using a string potentiometer (L Series, FirstMark Controls) that was attached to the extremity of the piston (Figure 5). Data were recorded at a rate of 20 kHz with a Dewetron data acquisition system (DEWE-2010 Series, Dewetron, Graz, Austria). In post-processing the force, moment, and displacement data were debiased at time zero, and then filtered consistent with the methods used in previous costal cartilage experiment studies (Forman et al 2010; forces and moments filtered using SAE J211 Channel Filter Class 30, displacements filtered using CFC 1000).

The three-dimensional displacement of the sternum block was visualized stereoscopically via two high-speed digital imagers (Redlake, HG-100K, 1024 × 768 pixels, 8 bit (black and white)). These imagers were mounted on a 6 degrees-of-freedom adjustable beam. The imagers were angled approximately 20 degrees relative to each other, with a common focal point located approximately 150 cm in front of each of the imagers (approx. distance from the sensor, in-line with the view direction). The imagers were mounted upright, looking down at the anterior surface of the costal cartilage specimen. Images were collected concurrently at a frame rate of 1000 Hz (exposure 250 μs). Image collection timing was synced between the two cameras, and the sync signal was recorded by the data acquisition system. Lighting was provided by a self-cooling, continuous-intensity lighting system (Edmund optics, Dolan-Jenner MI-150 fiber optic illuminator equipped with a ½" flexible fiber optic light guide). The two-imager system was calibrated and the images were analyzed with the Aramis v6.2.0 software (GOM-Optical Measuring Techniques, Germany).

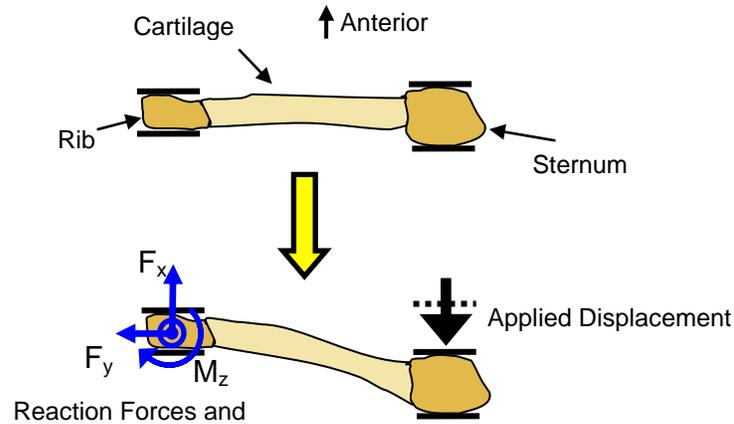

Figure 4: Schematic illustration of the boundary conditions that were used in the testing. The sternum end of the specimen was displaced posteriorly, with all other boundary conditions remaining fixed. Reaction forces and moments were measured at the rib ends of the specimens (the arrows shown indicate the positive force directions).

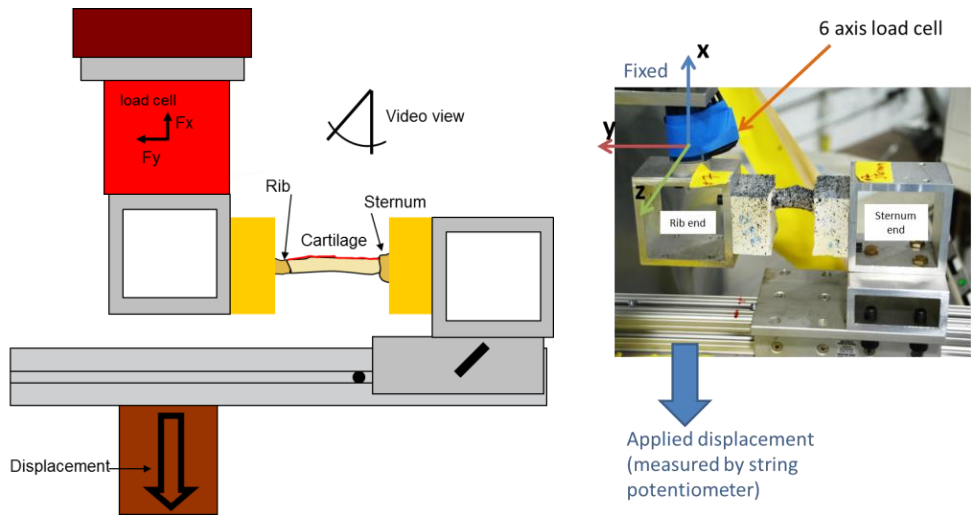

Figure 5: Schematic illustration and picture of the costal cartilage test setup.

## *Test methodology*

Prior to testing, each of the specimens was soaked in body temperature (37º C) saline solution for 2 hours. After this soaking time, each specimen was removed from the soaking solution and the surface was patted dry. The anterior surface of the specimen was then painted to prepare for visual strain capture and analysis. A base layer of white paint was applied via spray. An over-layer of a black speckle pattern was then applied via a combination of paint spray and manual application via felt marker.

### *Whole ribcage radiological images*

Radiological images of the whole ribcage were acquired for the intact subject, prior to performing any tests. Computed tomography was used (slice thickness: 0.625 mm, in-plane resolution: 0.98 × 0.98 mm).

### *Cartilage Calcification*

The magnitude of calcification in each of the costal cartilage specimens and in the whole chest was qualitatively assessed via CT scan (individual scans for the prepared specimens and the whole body scan for the whole chest). The degree of calcification was assessed using the scoring method developed by McCormick et al. (1980). That score is based on an ordinal scale ranging from 0 (no calcification) to 4.0 (very severe calcification), with increments of 0.5. For each specimen, and for the whole chest, the score was assessed via a visual examination of the CT scan from an anterior viewpoint, thresholded for semi-transparent visualization of the costal cartilage to allow examination of the interstitial calcification. Scores were based on a comparison to reference radiographs provided in McCormick *et al.* (1980).

### *Estimate of the CC young's modulus*

Forman (2009) described a relationship between the qualitative calcification score (McCormick *et al.,* 1980) and the effective material properties that would be required to model the costal cartilage as a homogeneous structure (incorporating the effects of the perichondrium into the homogeneous material representation). Forman developed a series of finite element models of the costal cartilage with various types and amount of calcification obtained from the samples he tested. Based on an inverse modeling approach (using the finite element method), the relationship between level of calcification in the costal cartilage and its material properties was established. Using the calcification scores determined from examinations of the CT scans, it was thus possible to estimate the effective pseudo-elastic moduli that would be required to model these specimens as homogeneous structures.

# Results

The costal cartilage segments from the right $2^{nd}$ rib through the $7^{th}$ rib were examined for consideration for testing. Of these, only the costal cartilage from ribs 4, 6, and 7 were deemed suitable for testing – all others exhibited fractures either in the cartilage or at the costo-chondral or costo-sternal junctions (which occurred during the table-top testing of this subject). Of the segments tested, only the $4^{th}$ rib segment consisted of the whole length of cartilage – the other two consisted of partial segments of the cartilage due to fractures at the extremities. For the partial segments, the terminal ends of the cartilage were potted directly into the FastCast® (instead of potting a portion of the attached rib or sternum).

### *Relative displacement of the CC extremities*

The three-dimensional displacement of the sternum block relative to the rib block was measured based on the high-speed images, by stereo-correlation. Because of the compliance of the CC, and the use of a cantilever beam (Figure 5) to allow the cameras to see the superior aspect of the CC samples, oscillations in the y and z directions were observed in the displacement of the sternum (Figure 6). The sternum block displacement is expressed in the same coordinate system as the load cell (Figure 5). Because of the oscillations of the cantilever beam, displacements in the positive y direction that led to the compression of the sample were observed.

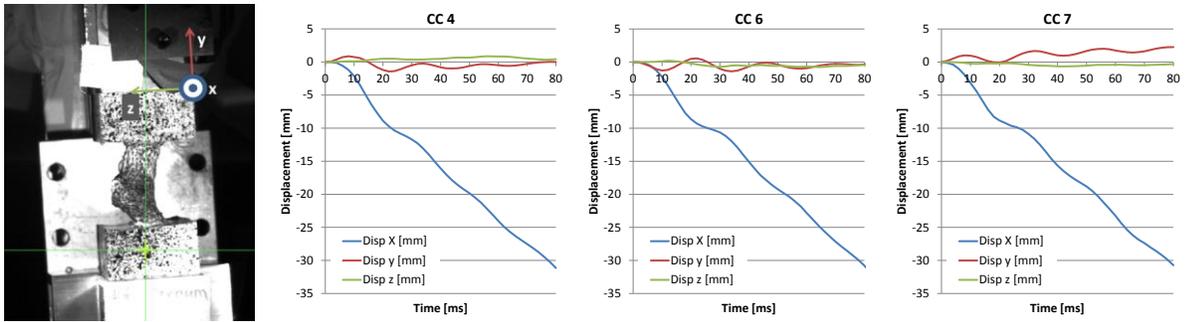

Figure 6: *Left:* Coordinate system defined to document the kinematics of the sternum block. This coordinate system is consistent with the load cell orientation (Figure 5). *Right*: measured 3D displacements of the sternum block relative to the rib block.

## *Force time-histories and fracture patterns*

The forces in the directions X, Y and Z were plotted against time for all the specimens (Figure 7, Figure 8, and Figure 9).

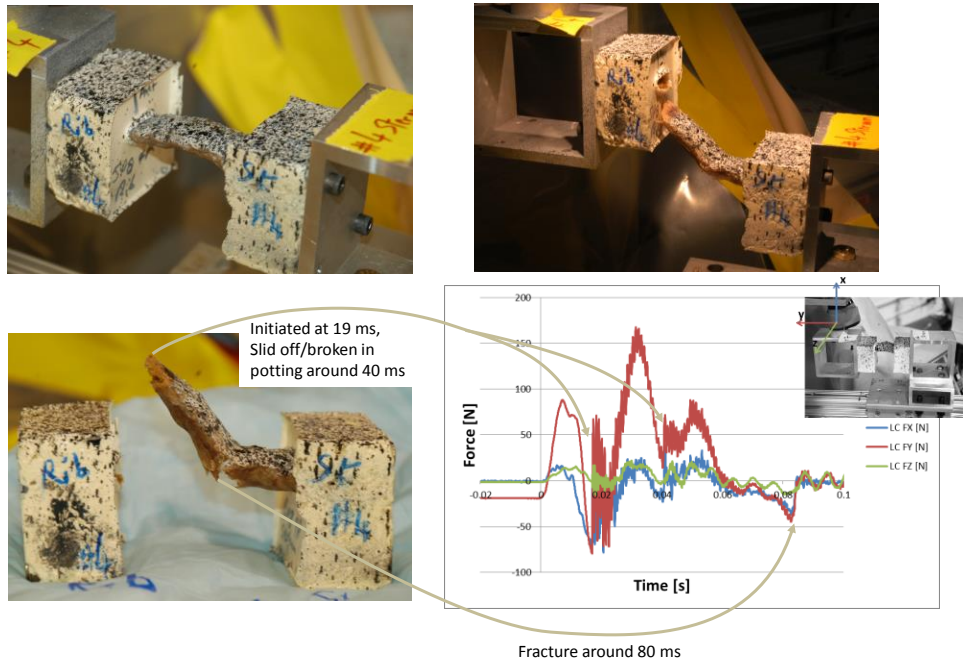

Figure 7: Overview of the mechanical response of CC4. It broke nearby the costo-sternal junction, and came off/broke in the potting on the rib side.

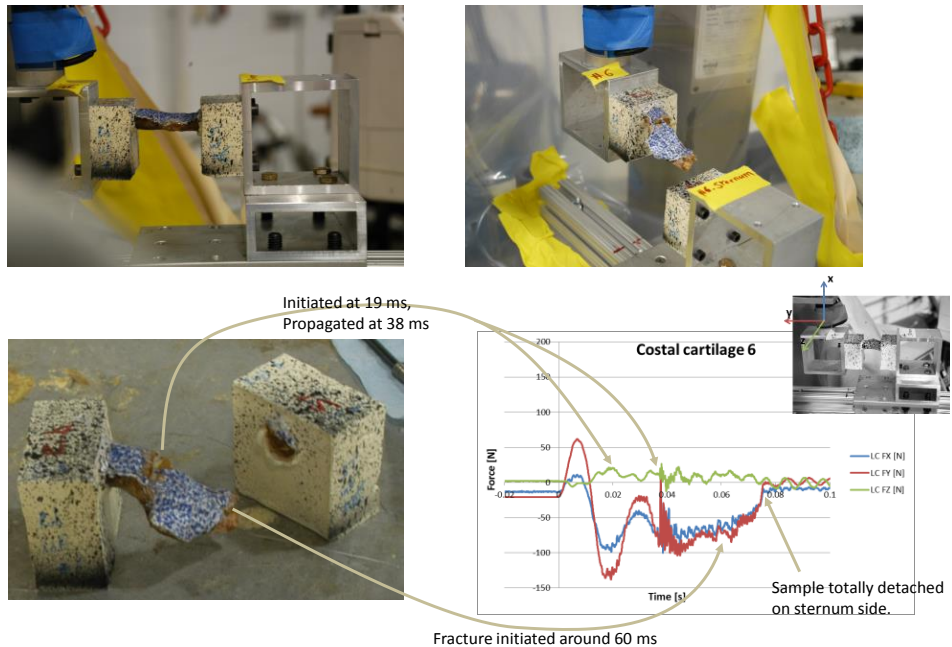

Figure 8: Overview of the mechanical response of CC6. It fractured in the cartilage on the rib side, and in the potting on the sternum side.

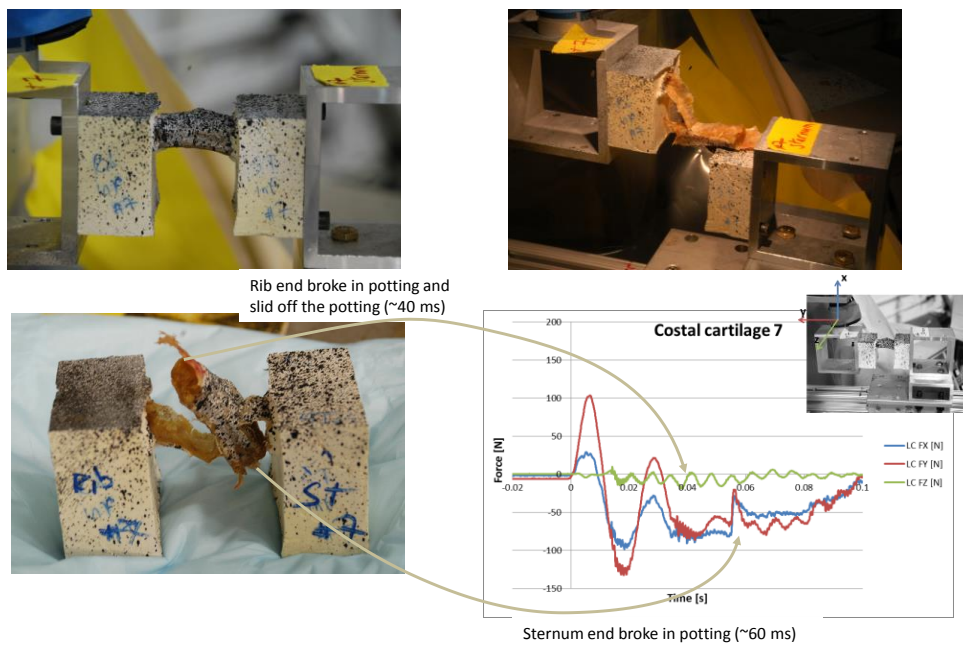

Figure 9: Overview of the mechanical response of CC7. It broke at both ends; the superior part of the perichondrium stayed attached to sternum end, and the inferior part of the perichondrium stayed attached to rib end.

## *Cartilage calcification*

The resulting calcification scores are shown in Table 1. The CT scan views that were used in this assessment are shown with the reference radiographs from McCormick *et al.* (1980) that were qualitatively deemed to most closely match the degree of calcification in each specimen. The effective Young's moduli were obtained relationship presented in Forman (2009).

Table 1: Qualitative calcification scores and estimated effective pseudo-elastic moduli.

| Specimen | Current study | McCormick et al. (1980) reference radiograph | Score | Eff. Modulus (MPa) |
|---|---|---|---|---|
| 4R* | 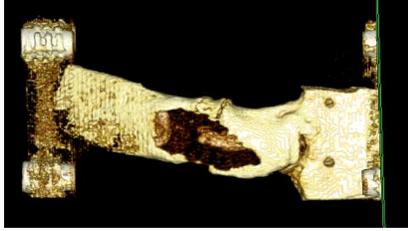 | 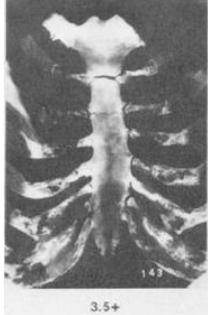 | 3.5 | 69 |
| 6R | 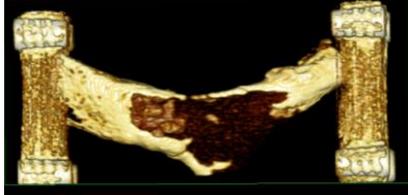 | 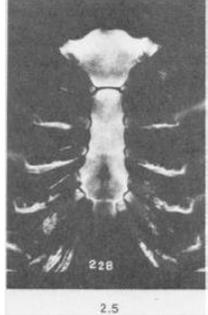 | 2.5 | 40 |
| 7R* | 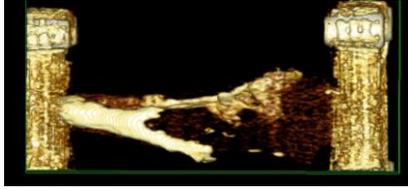 | 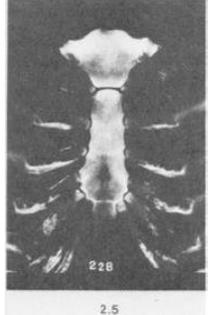 | 2.5 | 40 |
| Whole chest | 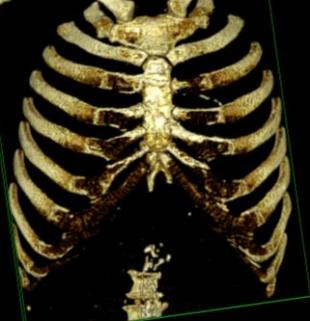 | 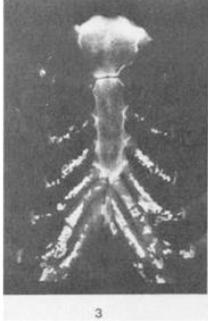 | 3.0 | 51 |

* Partial specimen

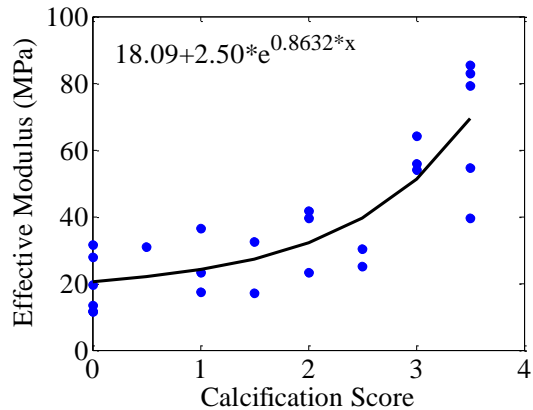

Figure 10: Estimated relationship between qualitative calcification score and the effective pseudo-elastic modulus required to model a costal cartilage segment as a homogeneous structure (reproduced from Forman 2009).

# Discussion

## *Oscillation in the sternum block displacement*

Figure 11 represents the position of the sternum mount at around 10 ms. The sternum mount is lower that the rib mount, and has moved towards the rib mount. As a result, a compression force that is directed upward is generated (the x and y components are both positive). Next, with the rapid increase of the displacement in the x-direction, the forces in the x- and y-direction decrease (more tension is applied to the sample), until the sternum mount moves back towards the rib mount because of its own oscillations.

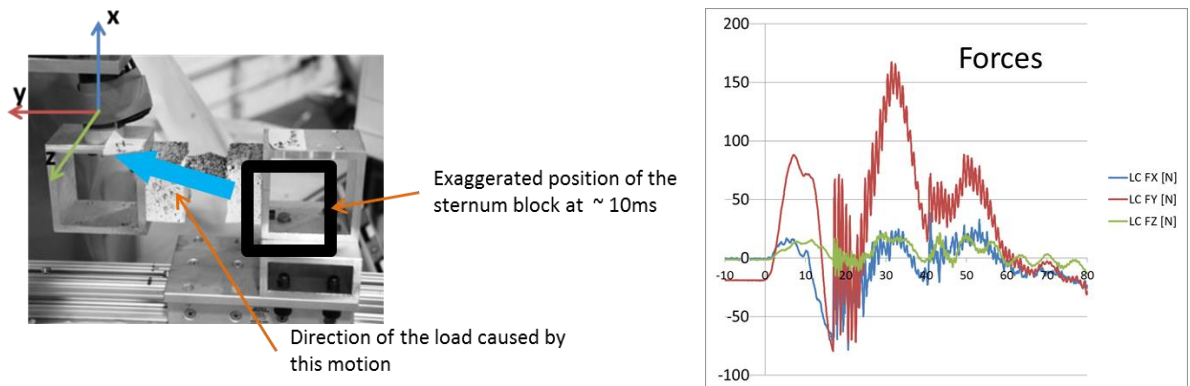

Figure 11: Effect of the displacement in the x and y directions on the force.

## *Estimate of the effective Young's moduli*

The effective material properties were determined based on a qualitative assessment of radiological images. The effective material properties – as used in this study – is the term used to describe material properties (specifically the elastic modulus) which are developed empirically so that a simplified, homogeneous finite element model of a costal cartilage segment (using the effective modulus) will match the overall structural behavior of a real costal cartilage segment. Effective properties are different from local material properties in that effective properties are developed empirically (e.g., through curve-fitting or inverse-FE) to produce a desired overall structural behavior, whereas local properties are measured locally through material property tests. Effective properties are useful when the goal of the model is to describe the overall mechanical response of the tissue, while not necessarily matching the local field of strain or stress.

## Conclusions

This paper reports the results of three dynamic shear tests performed on costal cartilage samples harvested from one PMHS. Because of the inertia of the sternum mount and the use of a cantilever beam, compression was applied to the samples. Although the motion of the sternum mount towards the rib mounts was minimal (less than 1.5 mm), the compression load was found to be in the same range as the shear load. This indicates that the costal cartilage does not behave like a soft tissue and that the amount of calcification has to be taken into account for the prediction of the costal cartilage response to dynamic loadings.

## *Appendix - Summary of the fracture reported in the right ribcage*

- Rib 1, (1) in the rib, posterior aspect, at the costo-vertebral junction (CVJ), and (2) at the sternochondral junction (SCJ),
- Rib 2, in the rib, posterior aspect, at the CVJ,
- Rib 4, in the rib, 15 mm posterior to costochondral junction (CCJ),
- Rib 5, (1) in the rib, 8 mm posterior to CCJ, and (2) at the SCJ,
- Rib 6, (1) in the rib, 20 mm posterior to CCJ, and (2) at the SCJ,
- Rib 7, (1) in the costal cartilage, 10 mm anterior to CCJ, and (2) at the SCJ.

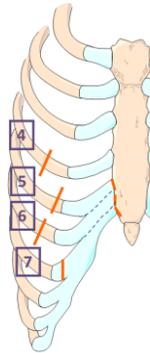

Figure 12: Rib and costal cartilage fractures in the right side of the ribcage.